\documentclass[lettersize,journal]{IEEEtran}
\usepackage{amsmath,amsfonts}
\usepackage{algorithmic}
\usepackage{algorithm}
\usepackage{array}
\usepackage[caption=false,font=tiny,labelfont=sf,textfont=sf]{subfig}
\usepackage{textcomp}
\usepackage{stfloats}
\usepackage{url}
\usepackage{verbatim}
\usepackage{graphicx}
\usepackage{cite}
\usepackage{multirow}
\usepackage{color}
\usepackage{makecell}
\usepackage{float}

\begin{document}

\title{Generative Model Watermarking Suppressing High-Frequency Artifacts}

\author{Li Zhang, Yong Liu, Xinpeng Zhang and Hanzhou Wu, \emph{Member, IEEE}
\thanks{\emph{Corresponding author: Hanzhou Wu (contact email: h.wu.phd@ieee.org)}}
}

\markboth{}%
{}


\maketitle

\begin{abstract}
Protecting deep neural networks (DNNs) against intellectual property (IP) infringement has attracted an increasing attention in recent years. Recent advances focus on IP protection of generative models, which embed the watermark information into the image generated by the model to be protected. Although the generated marked image has good visual quality, it introduces noticeable artifacts to the marked image in high-frequency area, which severely impairs the imperceptibility of the watermark and thereby reduces the security of the watermarking system. To deal with this problem, in this paper, we propose a novel framework for generative model watermarking that can suppress those high-frequency artifacts. The main idea of the proposed framework is to design a new watermark embedding network that can suppress high-frequency artifacts by applying anti-aliasing. To realize anti-aliasing, we use low-pass filtering for the internal sampling layers of the new watermark embedding network. Meanwhile, joint loss optimization and adversarial training are applied to enhance the effectiveness and robustness. Experimental results indicate that the marked model not only maintains the performance very well on the original task, but also demonstrates better imperceptibility and robustness on the watermarking task. This work reveals the importance of suppressing high-frequency artifacts for enhancing imperceptibility and security of generative model watermarking.
\end{abstract}


\section{Introduction}
\IEEEPARstart{T}{he} past decade has witnessed the success of deep learning \cite{lecun2015deep} in various fields such as computer vision \cite{krizhevsky2017imagenet, szegedy2015going, ronneberger2015u}, natural language processing \cite{hinton2012deep,morgan2011deep,sennrich2015neural} and games \cite{silver2017mastering}. As a very popular class of deep learning models, deep neural networks (DNNs) have been increasingly applied by technology companies in their consumer products so as to provide better service and increase profits. However, creating advanced DNN models needs lots of data, expert knowledge and powerful computing resources. Besides, due to the openness or semi-openness, it is possible for an unauthorized user to tamper or redistribute DNN models, which violates the intellectual property (IP) of these models as a kind of expensive digital asset. Therefore, we urgently need to find solutions to protect the IP of DNNs.

Fortunately, increasing methods \cite{uchida2017embedding,wang2020watermarking,fan2019rethinking,adi2018turning,zhang2018protecting,zhao2021watermarking,zhao2021structural,wu2020watermarking} are proposed in recent years to protect the IP of DNN models by applying digital watermarking \cite{cox2007digital} which embeds a message, typically called watermark, into a host signal by slightly modifying the signal. By extracting the embedded watermark from a target marked signal, we are capable of verifying the ownership of the signal. Because the marked signal may be intentionally attacked prior to watermark extraction, the watermark embedding procedure is required to be robust against attacks for reliable ownership identification. A straightforward idea to protect the IP of DNN models is to directly extend advanced watermarking strategies suited to media signals to the DNN models since media watermarking has been widely studied in the past two decades \cite{liu2016blind,chaabane2015qr,li2021spread}. However, unlike media signals that are static data, DNN models are functional. Simply watermarking DNN models in a way similar to media watermarking could significantly degrade the functionality of DNN models. It indicates that we urgently need to develop watermarking methods specifically for DNN models, which is referred to as DNN (model) watermarking. 

A DNN consists of three important parts, i.e., input, internal network, output. The most intuitive strategy to mark a DNN is modifying the internal network parameters of the model. Many existing methods are proposed along this direction such as \cite{uchida2017embedding,wang2020watermarking,fan2019rethinking}, \cite{chen2019deepmarks,wang2021riga,tartaglione2021delving,wang2022protecting}. In addition to parametric modification, hiding watermark information into the internal network structure has also been investigated. For example, Zhao \emph{et al.} \cite{zhao2021structural} propose a novel technique to embed the secret watermark in the network structure during the process of channel pruning in order that the pruned model not only performs very well on its original task, but also carries a secret watermark that can resist against all parameter-based attacks. Since a DNN possesses the ability to accomplish a specific task, we can also use the functionality of the DNN model for watermark embedding and watermark verification. Along this direction, many methods mark a DNN by fine-tuning the DNN with a set of carefully crafted samples called trigger samples. The embedded watermark is generally retrieved by interacting with the target model and analyzing the prediction results of the model on a set of trigger samples. Some related works can be found in \cite{adi2018turning,zhang2018protecting,zhao2021watermarking}, \cite{lin2022verifying}, \cite{wu2016separable}.

Unlike many existing methods that use the internal network weights/structures or classification labels of the trigger set to embed watermarks, recent studies apply model watermarking to generative networks. For example, Wu \emph{et al.} \cite{wu2020watermarking} propose a generative model watermarking framework for protecting generative networks, where any image generated by the host network to be protected is marked. By detecting the watermark in the images generated by the marked network, the method can be used to identify the ownership of the target model and find whether an image is generated by a certain model or not. Zhang \emph{et al.} \cite{zhang2020model,zhang2021deep} attempt to resist surrogate model attack by introducing a watermarking scheme for image processing networks. They embed an invisible watermark in the outputted image. When the attacker trains a surrogate model by using the input-output pairs containing the watermark, the watermark is also learned by the surrogate model. Thus, the watermark can be extracted from the outputted image of the surrogate model for ownership verification. Recently, Zhang \emph{et al.} \cite{zhang2022generative} take into account the influence of embedding data in different channels of the marked image. By exploiting the characteristics of human visual system, they design a generative model watermarking scheme with better visual quality of the image.

For evaluating a watermarking system, imperceptibility requires that the embedded watermark should not be perceived by any attacker \cite{guo2018watermarking,li2019piracy,li2019prove}. If an attacker cannot perceive the embedded watermark, he may not take action to attack the watermark. It implies that better imperceptibility can reduce the probability of the watermark to be attacked to a certain extent. Unlike watermarking for classification tasks, generative model watermarking uses the generated image, rather than the weights or structure of the host network or the trigger set, for ownership verification. The present methods for generative models \cite{wu2020watermarking}, \cite{zhang2020model,zhang2021deep,zhang2022generative} only consider the imperceptibility of the secret watermark in the spatial domain, i.e., embedding the watermark into the spatial domain by an imperceptible fashion, but none of them consider the imperceptibility in the frequency domain. As a result, they inevitably introduce artifacts in the frequency domain that degrade the imperceptibility, which will surely impair the security. Therefore, we urgently need to take action to suppress these frequency artifacts.

Unfortunately, the existing generative model watermarking methods \cite{wu2020watermarking}, \cite{zhang2020model,zhang2021deep,zhang2022generative} use a watermark embedding network, which is an independent network or the host network itself, to watermark the generated images. We find that there are many up-sampling and down-sampling operations in the watermark embedding network, which will cause high-frequency artifacts to appear in the marked image. Moreover, the watermark embedding network completes the watermarking task by adding perturbation (watermark) to the generated image, which may result in higher-degree high-frequency artifacts. Through analyzing the marked images generated by the aforementioned watermarking methods, we find that these marked images do have serious high-frequency artifacts in the frequency domain. Actually, even if the embedded watermark is invisible in the spatial domain, the presence of the watermark can be detected due to the lack of invisibility in the frequency domain. The poor imperceptibility in the frequency domain may make the IP protection invalid, which is an important problem to be solved. Therefore, in this paper, we propose to consider the imperceptibility of generative model watermarking in both the spatial domain and the frequency domain.

To enhance imperceptibility of generative model watermarking and solve the problems existed in the frequency domain, we propose a generative model watermarking framework that alleviates high-frequency artifacts. For this purpose, we exploit anti-aliasing for developing a watermark embedding network ensuring that the high-frequency artifacts of the marked images in the frequency domain can be alleviated. Experiments indicate that the proposed technique can not only guarantee the effectiveness of model watermarking, but also well alleviate the high-frequency artifacts. The proposed method reduces the detectability of the embedded watermark and makes IP protection more effective. In addition, the proposed method is easy to be expanded to different generative tasks and is robust against pre-processing and surrogate model attack, which demonstrate the superiority and potential of the proposed work.

The contributions of this paper are summarized as follows:

\begin{itemize}
\item{We are the first to propose that imperceptibility in generative model watermarking needs to be considered from both the spatial domain and the frequency domain. For the first time, we find high-frequency artifacts in the marked images, which demonstrates the lack of invisibility in the frequency domain for generative model watermarking.}
\item{To improve the imperceptibility of the generative model watermarking in the frequency domain, we propose a general watermarking framework to alleviate high-frequency artifacts. For this purpose, we propose to exploit the anti-aliasing strategy for developing a watermark embedding network that can suppress high-frequency artifacts.}
\item{Extensive experiments show that the proposed work alleviates the high-frequency artifacts and achieves better imperceptibility. Moreover, the proposed work is effective for different generative tasks and robust to pre-processing attacks and surrogate network attack.}
\end{itemize}

The rest structure of this paper is organized as follows. We firstly introduce necessary preliminary concepts in Section II, followed by the proposed method in Section III. Experimental results and analysis are then provided in Section IV. Finally, we conclude this work in Section V.

\begin{figure*}[!ht] 
  \centering
  \begin{minipage}[b]{1\linewidth} 
  \subfloat[]{
    \begin{minipage}[b]{0.15\linewidth} 
      \centering
          \includegraphics[width=\linewidth]{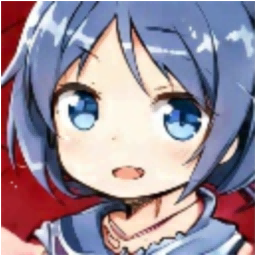}
    \end{minipage}
  }
  \hfill
   \subfloat[]{
    \begin{minipage}[b]{0.15\linewidth}
      \centering
      \includegraphics[width=\linewidth]{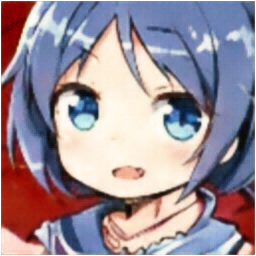}
    \end{minipage}
  }
  \hfill
    \subfloat[]{
    \begin{minipage}[b]{0.15\linewidth}
      \centering
      \includegraphics[width=\linewidth]{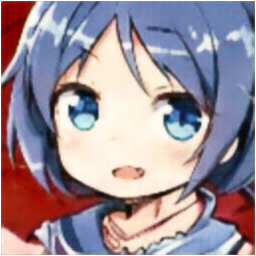}
    \end{minipage}
  }
  \hfill
    \subfloat[]{
    \begin{minipage}[b]{0.15\linewidth}
      \centering
      \includegraphics[width=\linewidth]{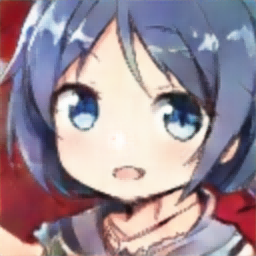}
    \end{minipage}
  }
  \hfill
    \subfloat[]{
    \begin{minipage}[b]{0.15\linewidth}
      \centering
      \includegraphics[width=\linewidth]{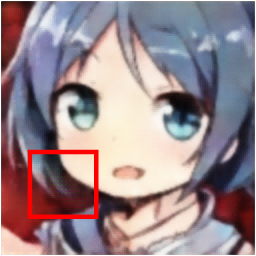}
    \end{minipage}
  }
  \hfill
    \subfloat[]{
    \begin{minipage}[b]{0.15\linewidth}
      \centering
      \includegraphics[width=\linewidth]{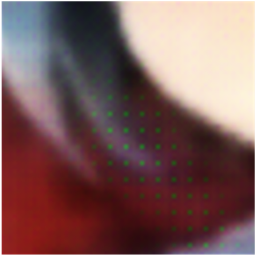}
    \end{minipage}
  }
  \end{minipage}
  \vfill
  \caption{An individual example of spatial image: (a) ground-truth image, (b) non-marked image generated by \cite{isola2017image}, (c) marked image generated by \cite{zhang2021deep}, (d) marked image generated by \cite{zhang2022generative}, (e) marked image generated by \cite{wu2020watermarking}, and (f) partial zoom-in of (e).}
\label{fig_1}
\end{figure*}

\begin{figure*}[!ht] 
  \centering
  \begin{minipage}[b]{1\linewidth} 
  \subfloat[]{
    \begin{minipage}[b]{0.17\linewidth} 
      \centering
          \includegraphics[width=\linewidth]{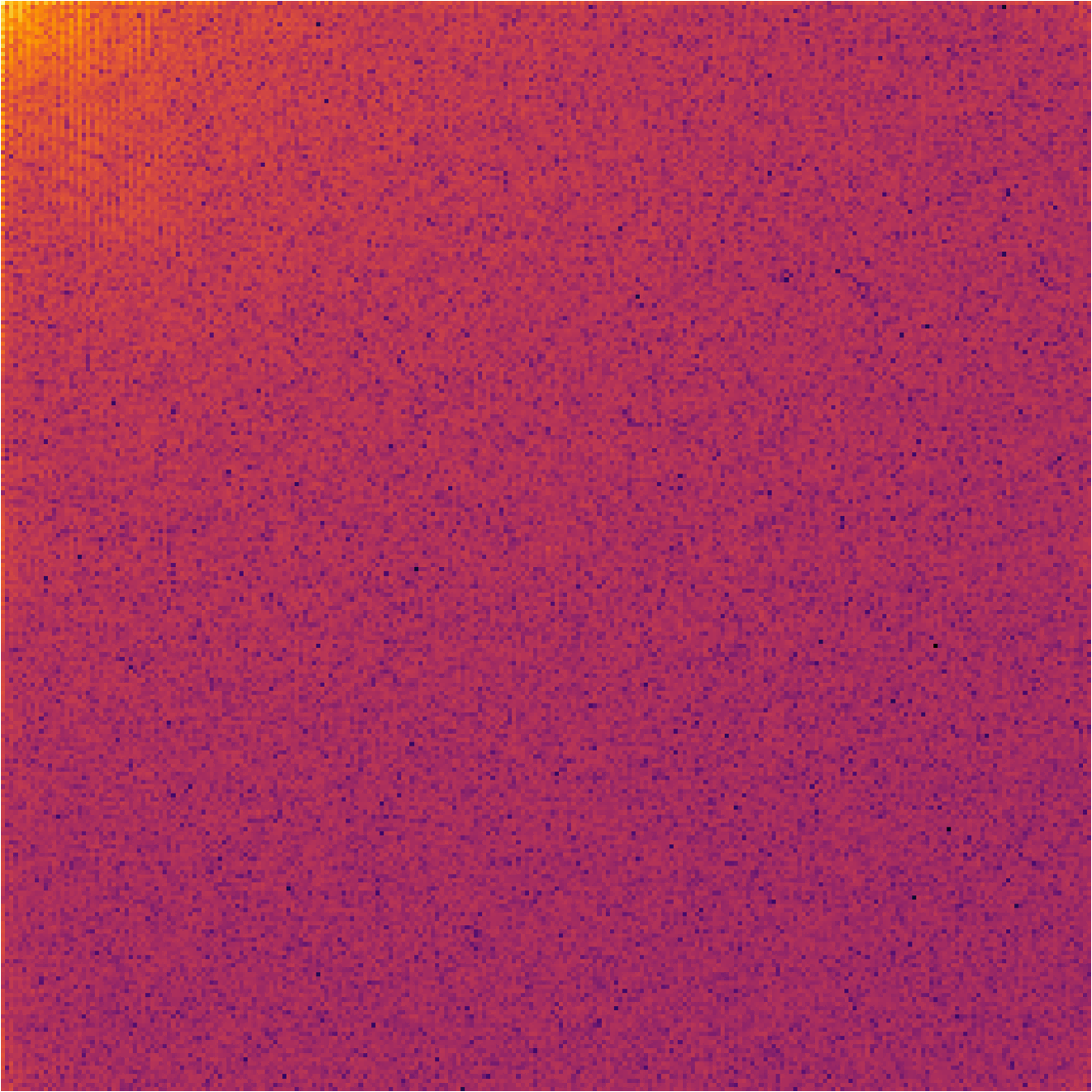}
    \end{minipage}
  }
  \hfill
   \subfloat[]{
    \begin{minipage}[b]{0.17\linewidth}
      \centering
      \includegraphics[width=\linewidth]{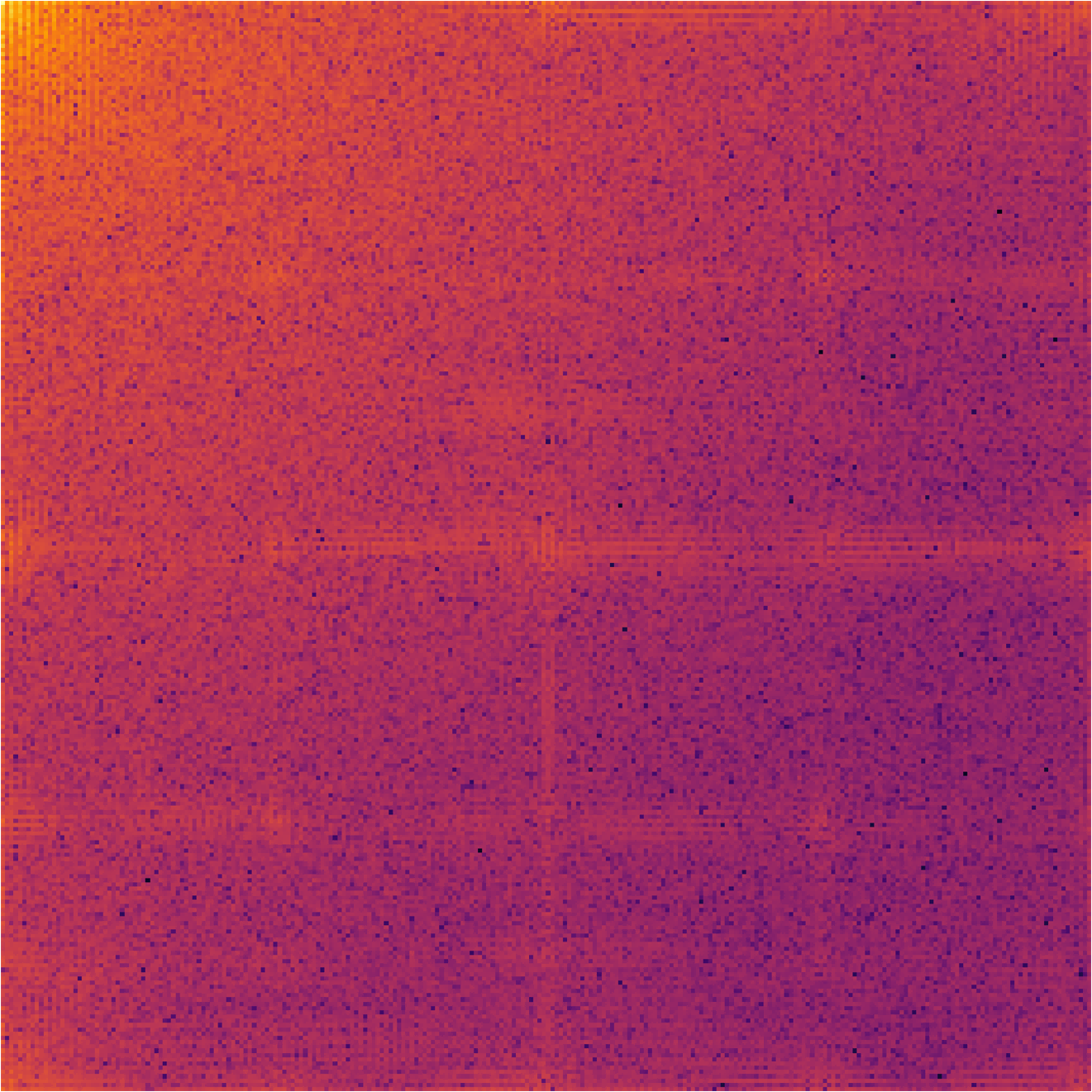}
    \end{minipage}
  }
  \hfill
    \subfloat[]{
    \begin{minipage}[b]{0.17\linewidth}
      \centering
      \includegraphics[width=\linewidth]{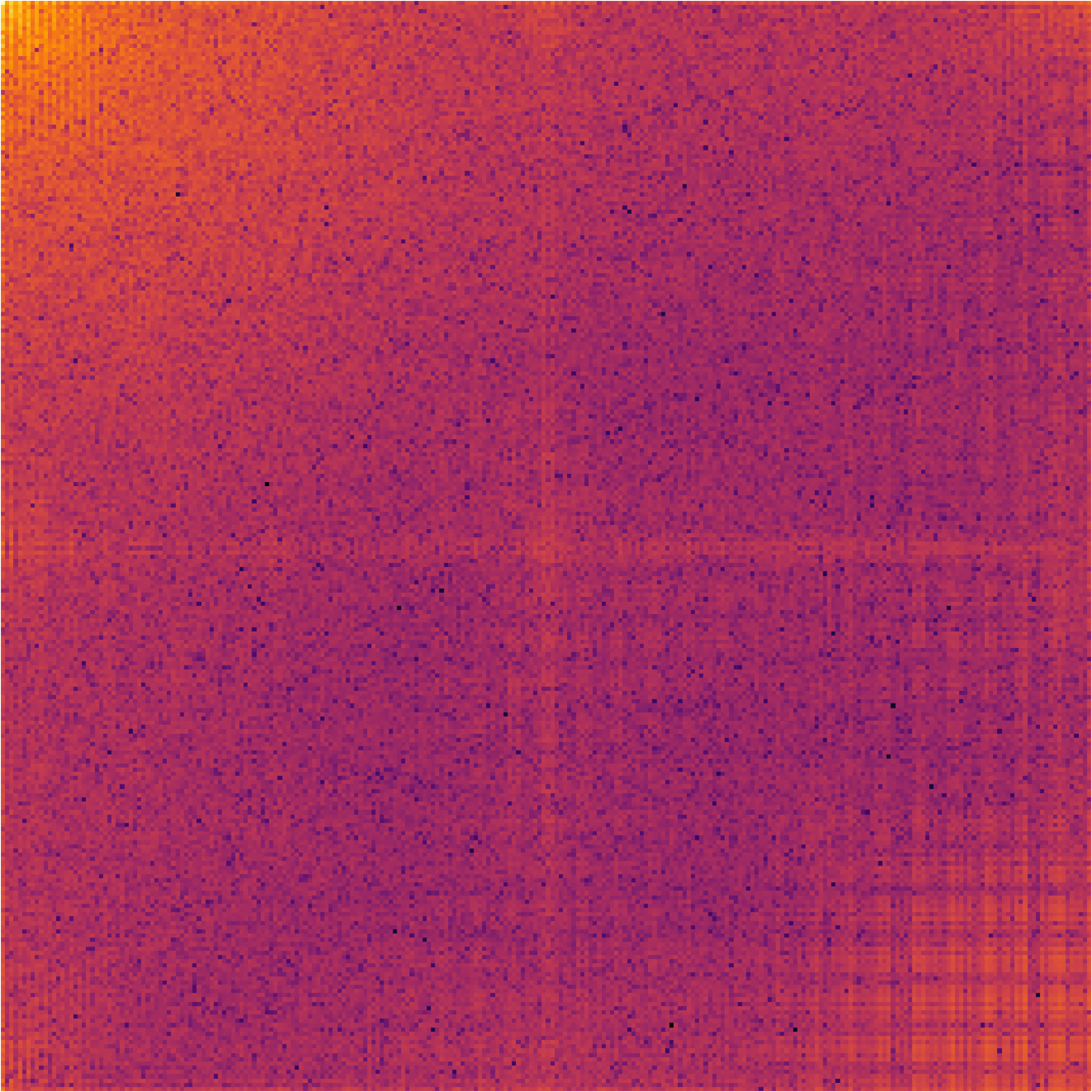}
    \end{minipage}
  }
  \hfill
    \subfloat[]{
    \begin{minipage}[b]{0.17\linewidth}
      \centering
      \includegraphics[width=\linewidth]{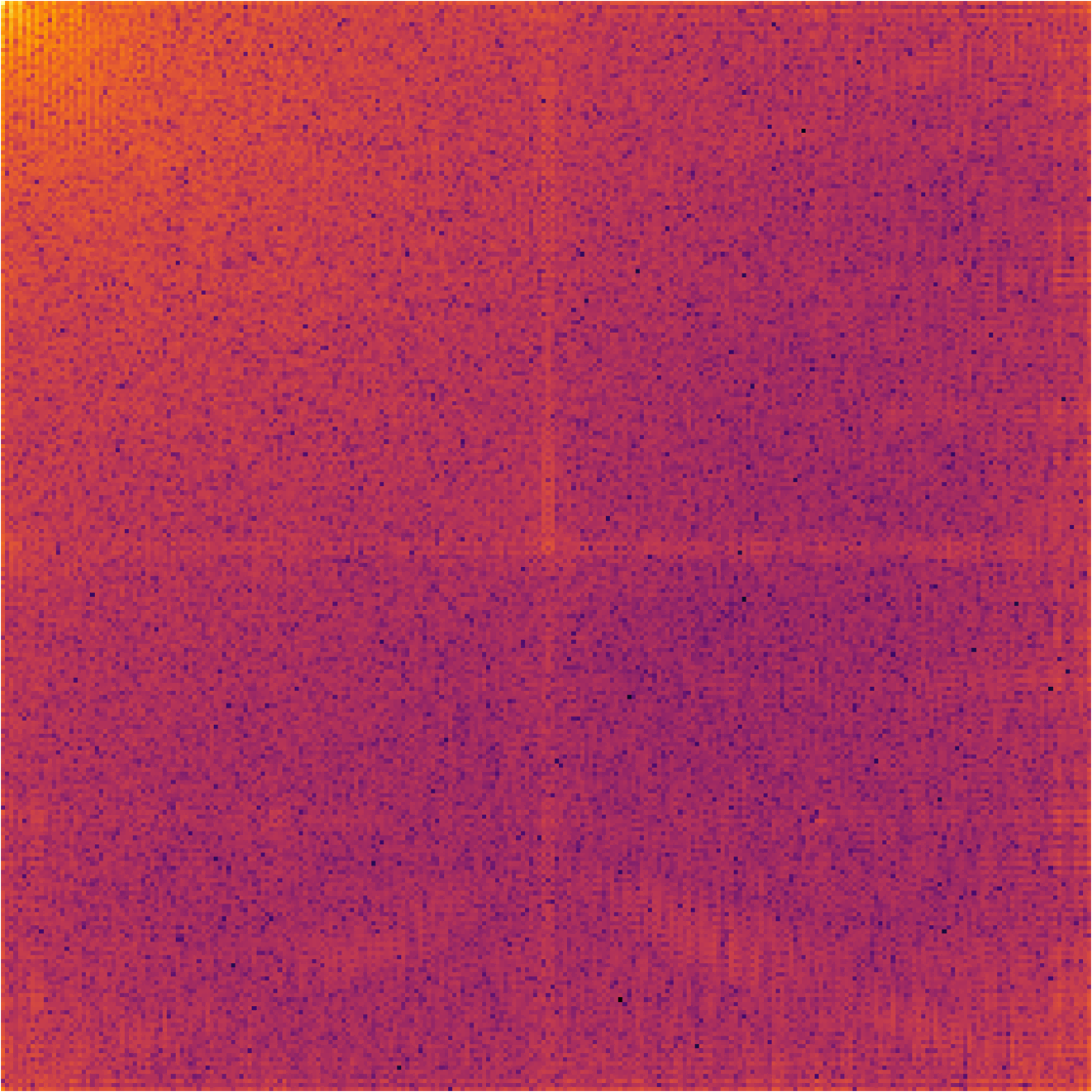}
    \end{minipage}
  }
  \hfill
    \subfloat[]{
    \begin{minipage}[b]{0.17\linewidth}
      \centering
      \includegraphics[width=\linewidth]{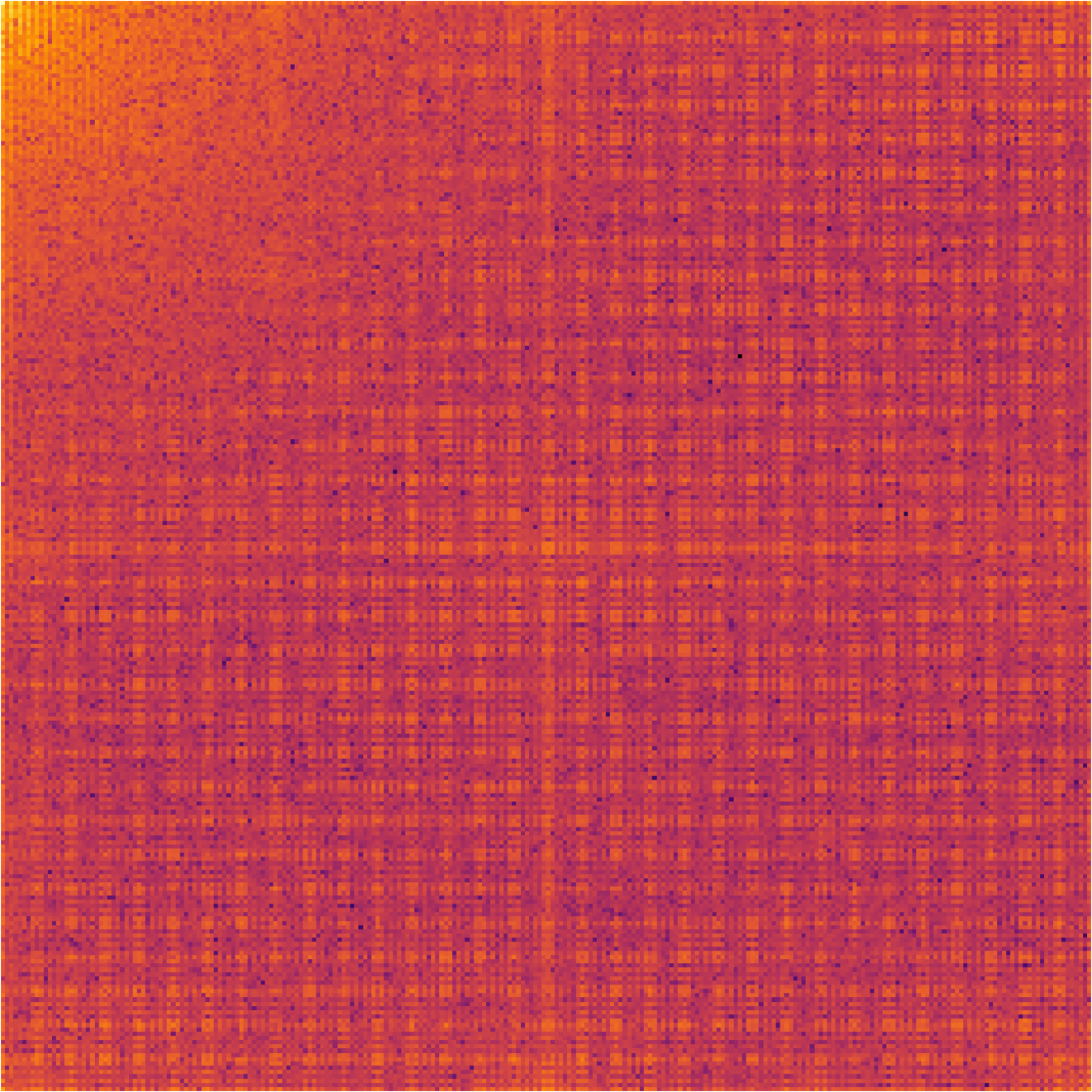}
    \end{minipage}
  }
   \hfill
    \begin{minipage}[b]{0.032\linewidth}
      \centering
      \includegraphics[width=\linewidth]{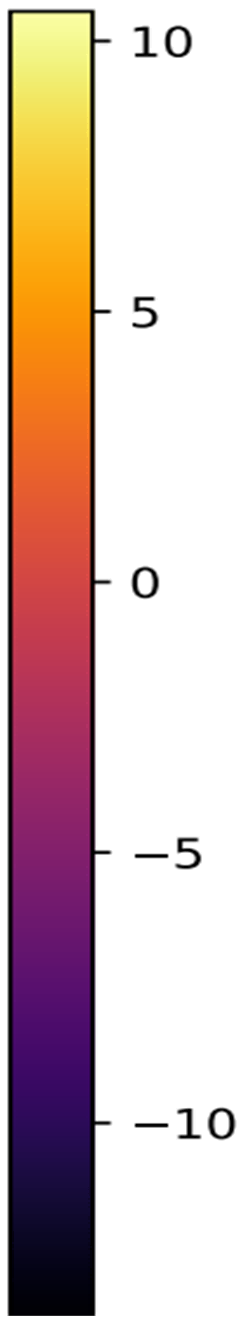}
    \end{minipage}
  \end{minipage}
  \vfill
  \caption{DCT heat maps over the corresponding test images (average results): (a) ground-truth images, (b) non-marked images generated by \cite{isola2017image}, (c) marked images generated by \cite{zhang2021deep}, (d) marked images generated by \cite{zhang2022generative}, (e) marked images generated by \cite{wu2020watermarking}.}
\label{fig_2}
\end{figure*}

\section{Preliminaries}
\subsection{Discrete Cosine Transform}
Discrete cosine transform (DCT) is a transform that converts a signal in the spatial domain to the frequency domain, which expresses a finite sequence as a sum of cosine functions with various frequencies. Suppose that we have a matrix $\textbf{A} \in \mathbb{R}^{n\times n}$ where $a_{i,j}$ represents the element at position $(i,j)$, after DCT transformation, we obtain the transformed matrix $\textbf{B} \in \mathbb{R}^{n\times n}$, where the element $b_{i,j}$ at position $(i,j)$ can be expressed as:
\begin{equation*}
b_{i,j} = c_ic_j\sum_{p=0}^{n-1}\sum_{q=0}^{n-1}a_{p,q} \text{cos}\left(\frac{(2p+1)i\pi}{2n}\right)\text{cos}\left(\frac{(2q+1)j\pi}{2n}\right),
\end{equation*}
where $c_i = c_j = 1/\sqrt{4n}$ for $i=j=0$, and $c_i = c_j = 1/\sqrt{2n}$ otherwise. $\textbf{A}$ can be reconstructed from $\textbf{B}$, i.e.,
\begin{equation*}
a_{i,j} = \sum_{p=0}^{n-1}\sum_{q=0}^{n-1}c_pc_qb_{p,q} \text{cos}\left(\frac{(2i+1)p\pi}{2n}\right)\text{cos}\left(\frac{(2j+1)q\pi}{2n}\right).
\end{equation*}

Previously, it has been demonstrated that images generated by generative adversarial networks (GANs) will exhibit severe artifacts in the frequency space, and these noticeable artifacts could be caused by up-sampling operations \cite{frank2020leveraging}. The existing generative model watermarking framework is GAN-like, i.e., there are many sampling operations in the watermark embedding network, which implies that the existing generative model watermarking frameworks are likely to introduce noticeable artifacts. Based on this key insight, we want to detect whether the marked images introduce frequency artifacts or not. 

We exploit DCT to detect whether the frequency artifacts exist or not. The DCT spectrum is plotted as a heat map, in which each point represents a coefficient of the corresponding frequency, and the larger the value at this point, the brighter it would look like \cite{frank2020leveraging}. The frequencies of the heat map increase from left to right and from up to bottom. So, the upper left region of the heat map corresponds to low frequencies and the lower right region corresponds to high frequencies. Due to the energy compression capability of the DCT, the magnitude of the coefficient decreases rapidly as the frequency increases.

\subsection{Detection of High-Frequency Artifacts}
To detect the frequency artifacts, we use the widely applied dataset Danbooru2019 \cite{branwen2019danbooru2019} for experiments. The original task of the host network is limited to paint transfer \cite{isola2017image}. And, three model watermarking methods introduced in \cite{zhang2021deep, zhang2022generative, wu2020watermarking} are tested. Fig. \ref{fig_1} gives an example. Fig. \ref{fig_1} (a) is the ground-truth image randomly selected from the Danbooru2019 dataset and Fig. \ref{fig_1} (b) is the non-marked image generated by  \cite{isola2017image}. Fig. \ref{fig_1} (c, d, e) are the marked images generated by \cite{zhang2021deep}, \cite{zhang2022generative}, \cite{wu2020watermarking}, respectively. The five images shown in Fig. \ref{fig_1} are visually close to each other at the first sight although Fig. \ref{fig_1} (a) shows better visual quality compared with other images since Fig. \ref{fig_1} (a) is the ground-truth image to be approximated.  This indicates that the original task can be successfully accomplished by \cite{isola2017image}, and \cite{wu2020watermarking}, \cite{zhang2021deep}, \cite{zhang2022generative} can all complete the watermarking task without severely compromising the original task. However, by careful observation, we still find slight spatial artifacts in the generated marked image, e.g., in Fig. \ref{fig_1} (e), we can find noticeable green points within the red-colored square, which have been enlarged in Fig. \ref{fig_1} (f). Spatial artifacts are anomalous states (e.g., green points, red lines and so on) in the spatial domain, which should be avoided in order to generate high-quality images.

We further analyze the characteristics of the test images in the frequency domain. Fig. \ref{fig_2} shows the DCT heat maps over the corresponding test images for different methods. Generally, for a natural image, the low frequency information is where the image intensity is smoothly transformed and represents most of the information of the image. In contrast, the edges in the image represent abrupt changes in pixels that should be approximated by the higher frequency functions. Therefore, natural images usually have most of their energy concentrated in the low-frequency part \cite{burton1987color, tolhurst1992amplitude}. The results given in Fig. \ref{fig_2} are in line with this conclusion. High-frequency artifacts are some unnatural and inconsistent textures in the frequency domain (e.g., grid-like patterns). In Fig. \ref{fig_2} (b), we can clearly observe the grid-like high-frequency artifacts. In Fig. \ref{fig_2} (c, d, e), we can easily find more serious grid-like high-frequency artifacts. Therefore, with the DCT domain detection results, we first find severe high-frequency artifacts appeared in generative model watermarking, which reveals the lack of the frequency imperceptibility in generative model watermarking. This may lead an attacker to perceive the embedded watermark through certain attacks and remove the watermark or claim ownership. We believe that the lack of frequency invisibility of generative model watermarking should be seriously considered. To enhance the invisibility, we are to analyze the reasons for the appearance of high-frequency artifacts in next subsection.

\subsection{Analysis of High-Frequency Artifacts}
Odena \emph{et al.} \cite{odena2016deconvolution} have linked grid-like artifacts in the spatial image to up-sampling. The authors in \cite{frank2020leveraging} also point that up-sampling used in a generative network forms a mapping from the low-dimensional latent space to the high-dimensional data space, which results in the appearance of inevitable grid-like high-frequency artifacts in the frequency domain. We believe that the marked image will introduce high-frequency artifacts due to many sampling operations in the watermark embedding network in generative model watermarking.

Furthermore, it is mentioned in \cite{zeng2021rethinking} that, when using triggers for image patching in a backdoor attack, high-frequency artifacts will be introduced directly because the triggers themselves may carry inherent high-frequency artifacts. In addition, since the addition of triggers leads to a decrease in correlation between adjacent pixels, a high-frequency function is needed to approximate the patched data. The authors in \cite{zhang2020udh} find that when embedding information using a DNN, the information will be added to the carrier image in the form of encoded high-frequency information. We argue that the process of embedding a watermark in generative model watermarking is similar to the backdoor attack or DNN-based information embedding. In generative model watermarking, when to embed the secret watermark into the generated image by the watermark embedding network, it is essentially adding a special perturbation to the marked image. Because of the time-frequency coherence, this directly leaves traces in the frequency domain, which leads to the creation of high-frequency artifacts.

In brief summary, we can assume that the presence of high-frequency artifacts in generative model watermarking comes from two main aspects. One is the structural defect of the watermark embedding network, and the other is the perturbation introduced by the process of embedding the watermark. The existence of high-frequency artifacts makes model watermarking exposed in the frequency domain, which allows attackers to easily discover the watermark in the frequency domain and thus remove it or declare the ownership. So, we urgently need to address the problem that generative model watermarking lacks invisibility in the frequency domain. It motivates us to design a novel watermark embedding network in this paper to alleviate the high-frequency artifacts.

\begin{figure*}[!t]
\centering
\includegraphics[width=\linewidth]{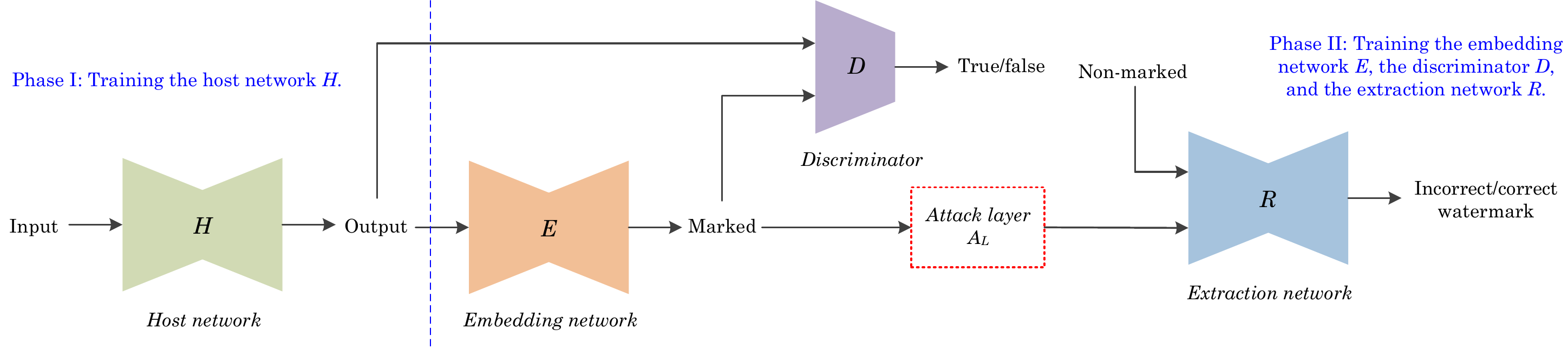}
\caption{Sketch for the proposed generative model watermarking framework which consists of the host network $H$, the embedding network $E$, the discriminator $D$, the extraction network $R$ and the attack layer $A_L$.}
\label{fig_3}
\end{figure*}

\section{Proposed Method}
The present generative model watermarking techniques can be divided into two categories. One is to use the host network to accomplish both the original task and the watermarking task simultaneously \cite{wu2020watermarking,zhang2022generative}, which implies that the host network has to serve as the role of watermark embedding. The other is to adopt a watermark embedding network independent of the host network to accomplish the watermarking task \cite{zhang2020model,zhang2021deep}. Compared with the former category, the latter one enables the process of model training to be faster, namely, the host model converges faster. To benefit from it, in this paper, we introduce an additional network for watermark embedding. Let $H$ be the host network to be protected. After the network $H$ completes the image processing or generation task, the outputted image is fed into the watermark embedding network $E$ to generate a marked image, which will be fed into a watermark extraction network $R$ for watermark retrieval to verify the ownership.

As pointed above, the present generative model watermarking techniques introduce noticeable artifacts in the frequency domain, which will surely degrade the imperceptibility of the embedded watermark. This is an urgent problem to be solved. Therefore, to improve the imperceptibility of generative model watermarking, in this paper, we propose a generative model watermarking framework that alleviates high-frequency artifacts. The proposed framework can obtain good concealment in the spatial domain and frequency domain while protecting the IP of the DNN well. In addition, considering the case that an attacker may attack the DNN even if he cannot perceive the embedded watermark, we also consider the robustness of the proposed method against common pre-processing attacks and surrogate network attack.

\subsection{General Framework}
Fig. \ref{fig_3} has provided the general framework. It is inferred that the proposed framework consists of five main modules, i.e., the host network $H$, the watermark embedding network $E$, the watermark extraction network $R$, the discriminator $D$ and the attack layer $A_L$. The original task of $H$ is to accomplish a task related to image processing or image generation such as paint transfer \cite{isola2017image}, style transfer \cite{wu2021hiding}, semantic segmentation \cite{isola2017image},  deraining \cite{yang2019joint}, and so on. Generally, $H$ accepts an image (or multiple images if necessary) as input and generates an image as output. For the sake of simplicity, we define the input space of $H$ as domain $S_\text{in}$ and the output space as domain $S_\text{out}$. 

We mark the images in $S_\text{out}$ so that by extracting watermarks from these images, we can protect the IP of $H$. In order to achieve this purpose, we apply the network $E$ to the images in $S_\text{out}$. The output space of $E$ is defined as domain $S_\text{marked}$. The images in $S_\text{marked}$ are marked. To not degrade the quality of the marked images, it is necessary that the images in $S_\text{out}$ and the images in $S_\text{marked}$ are visually consistent, which can be controlled by optimizing a loss function. To further improve the visual quality of the marked images, the discriminator $D$ is applied. It is noted that during the process of training $H$, the image generated by $H$ may also be fed into a discriminator, which depends on the original task of $H$ and is not the main interest of this paper. The marked images in $S_\text{marked}$ will be fed into the network $R$ for watermark extraction. To prevent $R$ from over-fitting, we feed both marked images and non-marked images into $R$ for watermark extraction. It is expected that only the marked images enable us to extract the correct watermark, while non-marked images reveal nothing about the watermark. To this purpose, we will force $R$ to output a noise image when a non-marked image was fed into $R$.

\begin{figure}[!t]
\centering
\includegraphics[width=\linewidth]{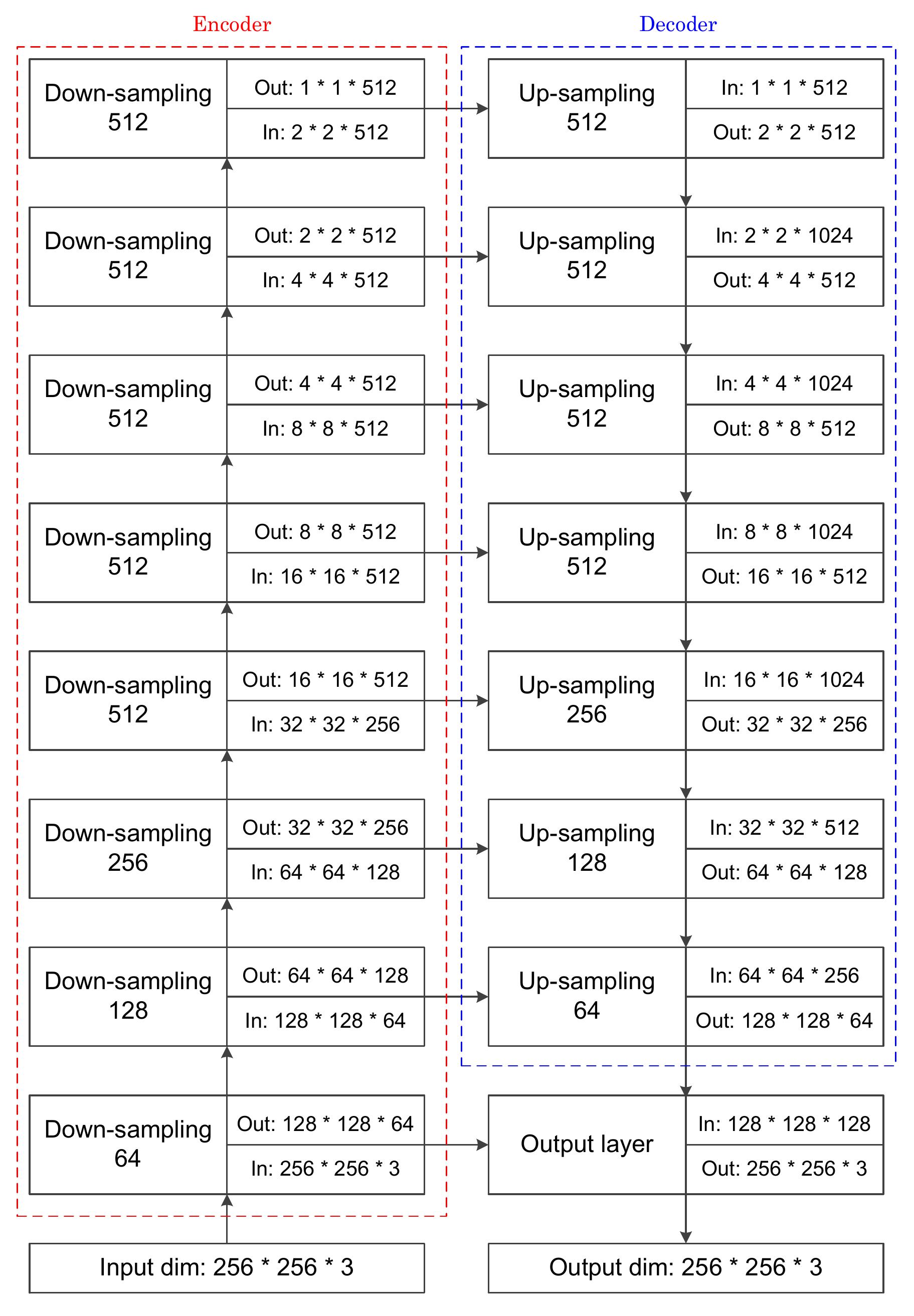}
\caption{The network structure for the watermark embedding network $E$.}
\label{fig_4}
\end{figure}

However, prior to watermark extraction, the marked images may be attacked. To enhance the proposed framework against common attacks, one of the most popular solutions is to mimic realistic attacks during model training. Therefore, we add the attack layer $A_L$ between $E$ and $R$ so that the watermark can be still extracted by $R$ even if the marked image was attacked. On the other hand, $H$ may be deployed in the form of application programming interface (API). It is possible that the attacker collects a number of input-output pairs by interacting with the API, which enable the attacker to train a surrogate model of $H$. Since the images generated by $H$ are all embedded with a watermark, in addition to the original input images, only the marked images may be collected by the attacker for training the surrogate model. To ensure that the proposed method can resist surrogate attack, we introduce an additional adversarial training stage. That is, we use two surrogate networks $N_1$ and $N_2$ to enhance the ability of the proposed framework against surrogate attack. $N_1$ aims to learn the mapping: $S_\text{in} \mapsto S_\text{marked}$, and $N_2$ aims to learn the mapping: $S_\text{in} \mapsto S_\text{out}$. We hope that $R$ can extract the correct watermark from the image generated by $N_1$, but extract a noise image from the image generated by $N_2$. To this purpose, after completing the initial training stage, we fix $E$ and fine-tune $R$. It has been demonstrated in \cite{zhang2021deep} that fine-tuning the watermark extraction network using a surrogate network can enhance the extraction ability, which can be well generalized to other surrogate networks.

\begin{figure}[!t]
\centering
\includegraphics[width=\linewidth]{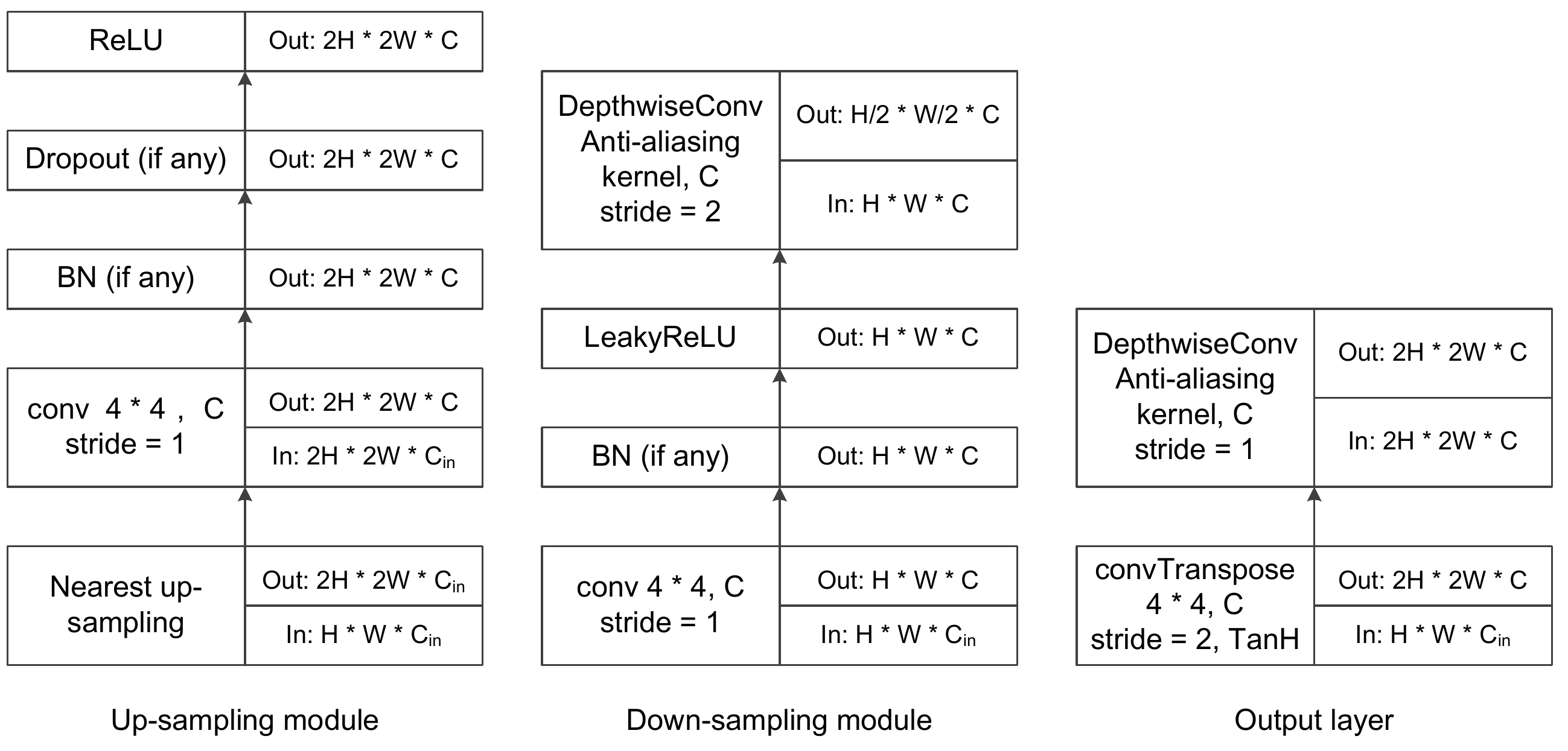}
\caption{Structural details for different functional modules. All the up-sampling and down-sampling layers except for the first down-sampling layer use BN. The first two up-sampling layers use dropout (= 0.5).}
\label{fig_5}
\end{figure}

\subsection{Structural Design}
As mentioned previously, there is no strict limitation to the architecture of the generative network $H$ to be protected. Any efficient backbone network may be applied to $H$ as long as it performs very well on the original task. It is also open for us to design the discriminator $D$. In this paper, we empirically use the popular network called PatchGAN \cite{isola2017image} to play the role of the discriminator due to its superior performance. U-net \cite{ronneberger2015u} has been widely applied in computer vision and proven to be effective for watermark extraction. Inspired by this, we use the U-net-like network in \cite{wu2021hiding} to act as $R$, but the input dimension and output dimension are changed to $256\times 256\times 3$. Meanwhile, to match the input and output, a pair of up-sampling and down-sampling layers are placed in the first layer and the last layer, respectively. $N_1$ and $N_2$ use the same architecture as $R$. 

How to design the network $E$ is the most important problem to be addressed. The goal of $E$ is to mark the image generated by the host network $H$. It is required that the visual distortion between the marked image and the input non-marked image should be kept low. Meanwhile, to guarantee the imperceptibility of the embedded watermark, the marked image should not introduce noticeable high-frequency artifacts. To this purpose, we design the network structure of $E$ as Fig. \ref{fig_4}. It was inspired by the popular U-net architecture \cite{ronneberger2015u}. However, different from U-net that applies max-pooling to capture region of interest for efficient image segmentation, we skip max-pooling because it will cause much useful information to be lost and thus impair watermark embedding. In Fig. \ref{fig_4}, the embedding network $E$ is consisting of an encoder, a decoder and an output layer. The encoder contains 8 down-sampling layers, whereas the decoder contains 7 up-sampling layers. The authors in \cite{zhang2019making} pointed out that aliasing can be mitigated by integrating low-pass filtering into the down-sampling process. From the viewpoint of model watermarking, since aliasing can be treated as a kind of high-frequency artifact, in order not to impair the image processing or generation task, it is feasible to suppress the high-frequency artifacts from the anti-aliasing perspective, which requires us to carefully develop the up-sampling module, down-sampling module and output layer shown in Fig. \ref{fig_4}. 

Motivated by the above insight, Fig. \ref{fig_5} provides the structural information for up-sampling module, down-sampling module and output layer. The idea for the structural design is described as follows. Traditional convolutional networks typically do not follow the sampling theorem, which will make them lack shift-invariant and therefore lead to aliasing. In signal processing, it is suggested to apply low-pass filtering before down-sampling to resist aliasing, which inspired Zhang \cite{zhang2019making} to perform low-pass filtering before down-sampling in network design. We use a similar approach for anti-aliasing. Specifically, for the down-sampling module, a $4\times 4$ convolutional kernel is firstly used for convolution, where the stride is set to 1. LeakyReLU \cite{xu2015empirical} is used as the activation function and batch normalization (BN) \cite{ioffe2015batch} is applied for accelerating convergence. More importantly, to suppress high-frequency artifacts and reduce computational overhead, we apply depth-wise convolution \cite{chollet2017xception}, in which the convolution kernel is fixed as a non-trained anti-aliasing kernel \cite{zhang2019making} to achieve low-pass filtering and the stride is set to 2 to achieve down-sampling. In this paper, the anti-aliasing kernel is empirically fixed as $(1,5,10,10,5,1)^\text{T}(1,5,10,10,5,1)$.

Up-sampling also requires anti-aliasing. However, in order to avoid excessive use of low-pass filters that result in overly smooth images, we hope to find another method to achieve up-sampling. Odena \emph{et al}. \cite{odena2016deconvolution} point that transpose convolution for up-sampling has nonuniform overlaps, which would cause artifacts. They find that connecting a convolutional layer after \emph{nearest (neighbor) up-sampling }\cite{gonzalez2009digital} can effectively solve this problem. This strategy has been applied in well-known GANs such as \cite{karras2017progressive}, \cite{karras2019style}. So, for the up-sampling module, we apply the traditional nearest up-sampling strategy, and the nearest up-sampling layer is followed by a convolutional layer. Moreover, BN and dropout \cite{srivastava2014dropout} are used for accelerating convergence and preventing over-fitting. The activation function is ReLU \cite{xu2015empirical}. 

We have found that high-frequency artifacts cannot be well suppressed if the output layer only uses an up-sampling layer. Therefore, we design the output layer to take the mirror of the down-sampling layer. Similarly, we apply a $4\times 4$ convolution kernel for transpose convolution (where the stride is set to 2) to achieve up-sampling, and connect a depth-wise convolution with a fixed convolution kernel for anti-aliasing. In this way, the high-frequency artifacts can be suppressed.

\subsection{Loss Function and Training Strategy}
The proposed framework includes two training phases. The first phase (called Phase I) is to train the host network $H$. The second phase (called Phase II) is to train the other networks. In other words, the two phases are independent of each other. The reason is that training all the networks together from scratch cannot ensure fast convergence of model training. However, it should be admitted that one may training the networks together if s/he can find an efficient joint-training strategy.

\subsubsection{Phase I} The host network $H$ will accomplish an image processing or generation task. Although the loss function relies on the task to be accomplished and can be very different from each other, it is generally required that the generated image is close to the ground-truth image. In order to improve the visual quality of the generated image, the adversarial loss function may be introduced. Nevertheless, in this paper, for the sake of simplicity, we use the neural network introduced in \cite{isola2017image} as $H$ and use the same training strategy described in \cite{isola2017image}.

\subsubsection{Phase II} The three neural networks $E$, $D$ and $R$ will be trained together from scratch. Basically, the watermark should be reliably embedded into the image generated by $H$, implying that we can use $R$ to extract the watermark from the marked image generated by $E$. It requires us to minimize
\begin{equation}
\mathcal{L}_1 = \frac{1}{|S_\text{in}|}\sum_{x\in S_\text{in}}|| R(E(H(x))) - w||_1,
\end{equation}
where $R(E(H(x)))$ represents the extracted watermark and $w$ means the ground-truth watermark. We will use $\ell_1$ norm as the distance measure by default unless otherwise specified. On the other hand, we hope that a noise-like image can be extracted from non-marked images, which inspires us to minimize
\begin{equation}
\mathcal{L}_2 = \frac{1}{2|S_\text{in}|}\sum_{x\in S_\text{in}}||R(x) - w_z||_1 + ||R(H(x)) - w_z||_1,
\end{equation}
where $w_z$ is a randomly generated image. So, the loss function for watermark extraction can be expressed as $\mathcal{L}_\text{ext} = \mathcal{L}_1 + \alpha\mathcal{L}_2$. where $\alpha$ is a tunable parameter. To embed the watermark in an invisible way, we minimize the visual distance between the image generated by $H$ and the image generated by $E$, which can be expressed as
\begin{equation}
\mathcal{L}_3 = \frac{1}{|S_\text{in}|}\sum_{x\in S_\text{in}}|| H(x) - E(H(x))||_1.
\end{equation}

Johnson \emph{et al.} \cite{johnson2016perceptual} encourage the use of a trained network to obtain the features of an image and thus calculate the distance between features for image quality assessment. Motivated by this suggestion, we use the features extracted by VGG19 \cite{simonyan2014very} for image quality assessment, i.e., we want to minimize
\begin{equation}
\begin{split}
& \mathcal{L}_4 = \frac{1}{5}\sum_{k=1}^{5}\mathcal{L}_{\text{VGG}_k}\\
& = \frac{1}{5}\sum_{k=1}^{5}\sum_{x\in S_\text{in}}\frac{||\text{VGG}_k(H(x))-\text{VGG}_k(E(H(x)))||_1}{|S_\text{in}|},
\end{split}
\end{equation}
where $\text{VGG}_k(\cdot)$ indicates the feature map extracted by the $k$-th layer of the pre-trained VGG19 model. The discriminator loss introduced in \cite{isola2017image} is used for $D$ in this paper to further improve the image quality. In this paper, ``false'' means marked images, whereas ``true'' means non-marked images. The discriminator loss is consisting of the true loss and the false loss, where the former is determined from the input image of $E$ and the ground-truth image (which is defined as the input image itself of $E$), and the false loss is computed from the input image of $E$ and the marked image generated by $E$. In other words, the discriminator loss can be expressed as
\begin{equation}
\begin{split}
\mathcal{L}_5 = \frac{1}{2|S_\text{in}|}\sum_{x\in S_\text{in}}& \text{log}(D[H(x), H(x)])\\
& + \text{log}(1-D[H(x), E(H(x))]).
\end{split}
\end{equation}

Therefore, the initial loss for training $E$, $D$ and $R$ is
\begin{equation}
\mathcal{L} = \mathcal{L}_\text{ext} + \beta_1\mathcal{L}_3 + \beta_2\mathcal{L}_4 + \beta_3\mathcal{L}_5,
\end{equation}
where $\beta_1$, $\beta_2$ and $\beta_3$ are tunable parameters. To improve the robustness, during model training, before inputting a marked image into $R$ for watermark extraction, $A_L$ can be applied to the marked image. This strategy has been demonstrated to be very effective in robustness enhancement of deep models. The next section will provide more experimental details about $A_L$. 

In addition, to enhance the ability of $R$ to resist the surrogate attack, we further introduce another adversarial training stage. During this training stage, we need to train two surrogate networks $N_1$, $N_2$ simultaneously. $N_1$ aims to learn the mapping: $S_\text{in} \mapsto S_\text{marked}$, and $N_2$ aims to learn the mapping: $S_\text{in} \mapsto S_\text{out}$. Therefore, we expect to optimize
\begin{equation}
\mathcal{L}_6 = \frac{1}{|S_\text{in}| }\sum_{x\in S_\text{in}}||  N_1(x) - E(H(x))||_1
\end{equation}
and
\begin{equation}
\mathcal{L}_7 = \frac{1}{| S_\text{in}| }\sum_{x\in S_\text{in}}||N_2(x) - H(x)||_1,
\end{equation}

After that, we want $R$ to extract the correct watermark from the output of $N_1$ and the noise-like image from the output of $N_2$. It requires us to minimize
\begin{equation}
\mathcal{L}_8 = \frac{1}{|S_\text{in}|}\sum_{x\in S_\text{in}}||R(N_1(x)) - w||_1 
\end{equation}
and
\begin{equation}
\mathcal{L}_9 = \frac{1}{|S_\text{in}|}\sum_{x\in S_\text{in}}||R(N_2(x)) - w_z||_1.
\end{equation}
In this stage, we keep $E$ unchanged and only fine-tune $R$.

\section{Experimental Results and Analysis}
In this section, we conduct extensive experiments and analysis for evaluating the performance of the proposed framework.

\subsection{Setup}
To verify the effectiveness, we evaluated the proposed work on two popular image processing tasks, i.e., paint transfer \cite{isola2017image} and style transfer \cite{wu2021hiding}. The popular Danbooru2019 dataset \cite{branwen2019danbooru2019} was used. A total of $4.5\times 10^4$ images were randomly selected for experiments. All these images were split into two disjoint subsets $S_A$ and $S_B$, where $|S_A| = 15,000$ and $|S_B| = 30,000$. $S_A$ and $S_B$ were respectively used for paint transfer and style transfer. The DNN in \cite{isola2017image} was used for paint transfer, and the DNN in \cite{wu2021hiding} was used for style transfer. For paint transfer, the method to create the images that belong to $S_\text{in}$ was the same as \cite{wu2020watermarking}. And, the method to create the ground-truth images for style transfer was the same as \cite{wu2021hiding}. For paint transfer, $8,000$ images in $S_A$ were used for Phase I, and the remaining $7,000$ images were split into four disjoint subsets used for Phase II. Specifically, $3,000$ images were used for training, $500$ images were used for validation, $500$ images were used for test, and $3,000$ images were used for the surrogate attack. Similarly, for style transfer, $15,000$ images in $S_B$ were used for Phase I, and the remaining images were split into four disjoint subsets for Phase II. To be specific, $6,000$ images were used for training, $1500$ images were used for validation, $1500$ images were used for test, and $6,000$ images were used for the surrogate attack. The resolution for all the images was set to $256\times 256\times 3$ by default. It is noted that the proposed work is not subjective to any particular image processing task, DNN or dataset. 

For Phase II, in the initial stage, $E$ and $R$ were trained from scratch by using $100$ epochs. In the adversarial training stage, the trained $R$ was fine-tuned with 50 epochs. The batch size was set to $1$. To test the surrogate attack, the surrogate network was trained with $200$ epochs, and the batch size was set to $8$. We empirically set $\alpha = \beta_1 = \beta_2 = 1$ and $\beta_3 = 0.001$. The Adam optimizer \cite{kingma2014adam} was used for training. The learning rate was set to $2.0 \times 10^{-4}$. Our implementation was executed on a single TITAN RTX 3090 GPU accelerated with CuDNN.

Due to the limited computational resource, the attack layer $A_L$ considered four common attacks including noise addition, resizing, JPEG compression and flipping. Specifically, for the noise addition attack, we applied the Gaussian noise with $\mu = 0$ and random $\sigma \in\left ( 0,0.2 \right ) $ to each input image for training. For the resizing attack, we resized the input image to a smaller or larger resolution in the range $[128^2, 512^2]$ by random and then resized the attacked image back to the original size for watermark extraction. For JPEG compression, we compressed the input images by applying a quality factor (QF) randomly selected from the range $[50, 90]$. For flipping, we flipped the image from left to right or from top to bottom by random. It is pointed that, although a limited number of common attacks were tested in this paper, it is always open for us to test more attacks as long as the computational resource is sufficient. 

Four common metrics were applied to evaluate the quality of the generated images, i.e., peak signal-to-noise ratio (PSNR), structural similarity (SSIM) \cite{wang2004image}, multi-scale structural similarity (MS-SSIM) \cite{wang2003multiscale}, visual information fidelity (VIF) \cite{sheikh2006image}. For each metric, the higher the value, the better the quality. Bit error rate (BER) was used to measure the reconstruction quality of binary watermark. It is determined as the percentage of incorrectly extracted bits when a binary stream was extracted. The lower the BER, the better the extraction performance. In addition, to evaluate the performance of watermark extraction, we defined a novel metric called success rate (SR). SR is determined as the percentage of successfully extracted watermarks. The watermark extraction process is deemed successful if the corresponding PSNR is higher than 35 dB or the corresponding BER is lower than $0.5\times 10^{-3}$. Finally, we used the result of DCT frequency domain detection to measure the frequency domain invisibility of model watermarking.

\begin{figure}[!t]
\centering
\includegraphics[width=\linewidth]{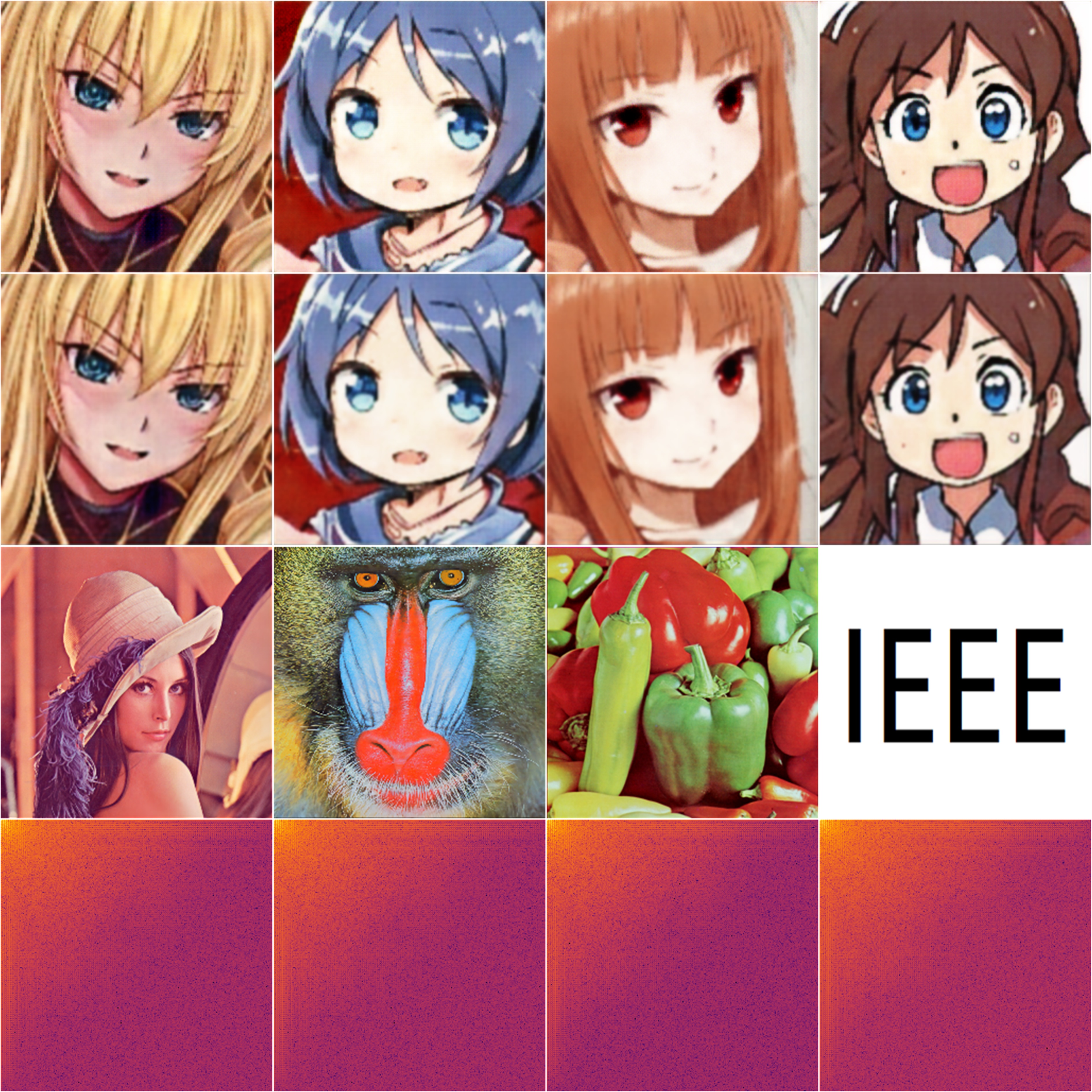}
\caption{Visual examples for paint transfer. Specifically, the images in the first row are non-marked. The images in the second row are marked. The images in the third row are the watermarks extracted from the corresponding marked images. The fourth row demonstrates the corresponding DCT heat maps.}
\label{fig_6}
\end{figure}

\subsection{Qualitative and Quantitative Results}
\begin{figure}[!t]
\centering
\includegraphics[width=\linewidth]{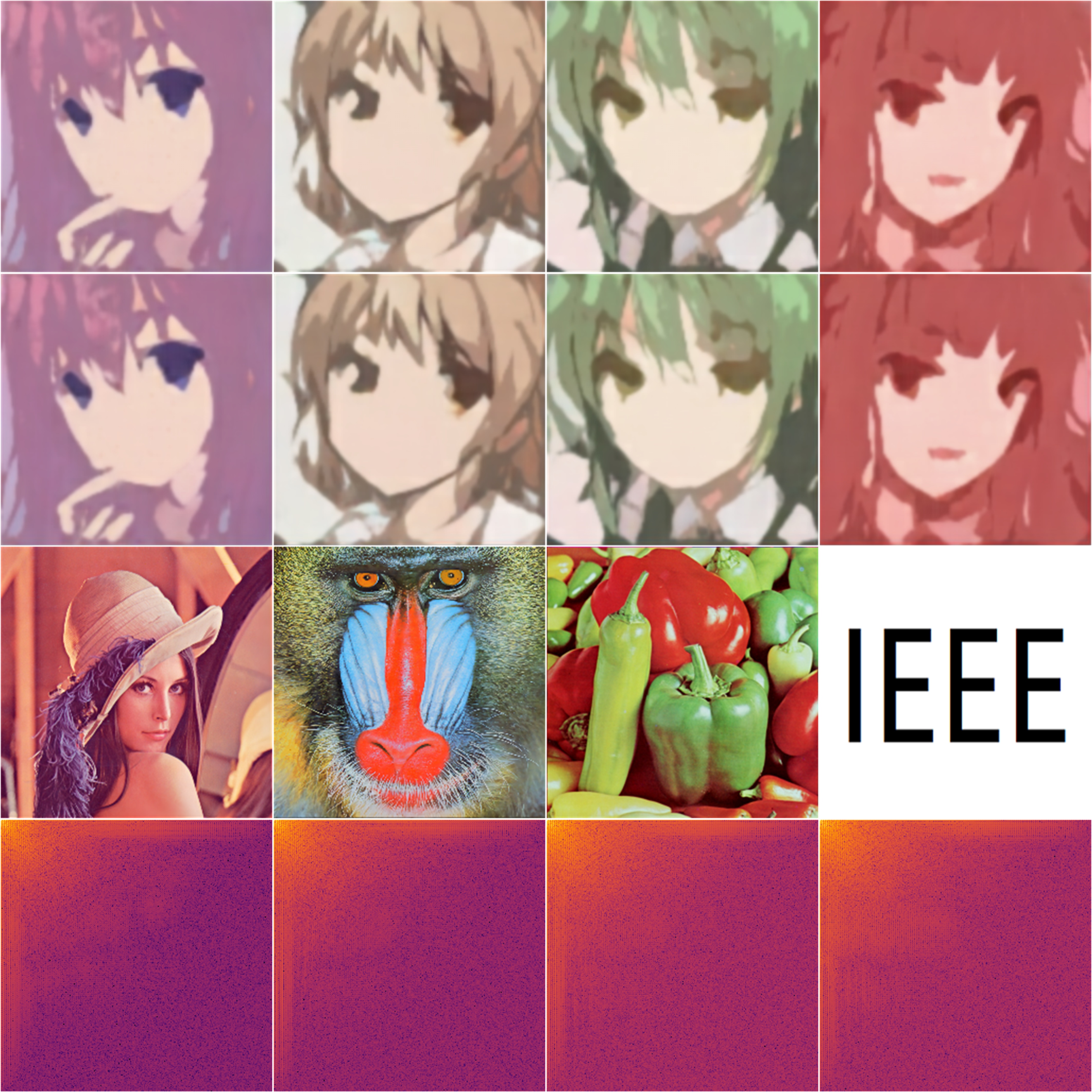}
\caption{Visual examples for style transfer. Specifically, the images in the first row are non-marked. The images in the second row are marked. The images in the third row are the watermarks extracted from the corresponding marked images. The fourth row demonstrates the corresponding DCT heat maps.}
\label{fig_7}
\end{figure}

We provide some qualitative and quantitative results in this subsection. We firstly provide some visual examples in Fig. \ref{fig_6} and Fig. \ref{fig_7}. Fig. \ref{fig_6} provides visual examples for paint transfer, whereas Fig. \ref{fig_7} provides visual examples for style transfer. It can be found that the marked images are visually close to the non-marked images, indicating that the proposed framework would not introduce noticeable artifacts in the spatial domain. It can also be found that the watermarks extracted from the corresponding marked images have high quality, meaning that the ownership of the marked model can be verified with high confidence. To further validate the superiority, we use PSNR, SSIM, MS-SSIM and VIF to measure the visual quality of the marked images over the test set. Meanwhile, we use PSNR, BER and SR to assess the quality of the extracted watermarks over the test set. The results are shown in Table \ref{tab:table1} and Table \ref{tab:table2}. The visual contents for the watermarks ``\emph{Lena}'', ``\emph{Baboon}'', ``\emph{Peppers}'' and ``\emph{IEEE}'' can be found in the third row of Fig. \ref{fig_6}. It is inferred from Table \ref{tab:table1} and Table \ref{tab:table2} that both the marked images and the extracted watermarks are of satisfactory quality, indicating that the proposed model watermarking framework achieves satisfactory imperceptibility in the spatial domain and achieves satisfactory performance on watermark extraction. It is worth mentioning that the SR is 100\% for all cases, which indicates that the extracted watermark can be used for reliable ownership verification of the target marked DNN model.

Yet another more important goal of this work is to suppress high-frequency artifacts. We analyze the characteristics of the test (marked) images in the frequency domain. Fig. \ref{fig_6} and Fig. \ref{fig_7} demonstrate the DCT heat maps over the corresponding test images for different watermarks. As shown in the fourth row of both figures, compared with Fig. \ref{fig_2} (b-e), the proposed framework removes the grid-like high-frequency artifacts caused by sampling and watermark embedding. It can be observed from Fig. \ref{fig_6} and Fig. \ref{fig_7} that the DCT heat maps are visually similar to Fig. \ref{fig_2} (a) which is corresponding to natural images. It indicates that the proposed generative model watermarking framework well alleviates the high-frequency artifacts, which ensures the invisibility of the embedded watermark.

\begin{table*}[!t]
\caption{Quality assessment for the marked images and the extracted color watermarks over the test set.  All experimental results shown in this Table are mean values. PSNR$_w$ measures the quality of the extracted color watermarks. \label{tab:table1}}
\centering
\begin{tabular}{c|c|cccc|c|c}
\hline\hline
Task & Watermark & Mean PSNR & Mean SSIM & Mean MS-SSIM & Mean VIF & Mean PSNR$_w$ & SR\\ 
\hline
Paint transfer & \emph{Lena} & 35.08 & 0.987 & 0.999 & 0.913 & 50.73 & 100\% \\ 
Paint transfer & \emph{Baboon} & 35.29 & 0.988 & 0.999 & 0.917 & 40.14 & 100\%  \\ 
Paint transfer & \emph{Peppers} & 35.02 & 0.986 & 0.999 & 0.914 & 44.97 & 100\%  \\ \hline
Style transfer & \emph{Lena} & 41.61 & 0.998 & 0.999 & 0.948 & 53.74 & 100\%  \\ 
Style transfer & \emph{Baboon} & 41.93 & 0.998 & 0.999 & 0.954 & 42.72 & 100\%  \\
Style transfer & \emph{Peppers} & 41.53 & 0.998 & 0.999 & 0.951 & 48.68 & 100\%  \\ \hline\hline
\end{tabular}
\end{table*}

\begin{table*}[!t]
\caption{Quality assessment for the marked images and the extracted binary watermarks over the test set.  All experimental results shown in this Table are mean values. BER measures the quality of the extracted binary watermarks. \label{tab:table2}}
\centering
\begin{tabular}{c|c|cccc|c|c}
\hline\hline
Task & Watermark & Mean PSNR & Mean SSIM & Mean MS-SSIM & Mean VIF & Mean BER & SR \\ \hline
Paint transfer & \emph{IEEE} & 34.98 & 0.986 & 0.999 & 0.913 & 0 & 100\% \\ \hline
Style transfer & \emph{IEEE} & 41.03 & 0.997 & 0.999 & 0.948 & 0 & 100\% \\ \hline\hline
\end{tabular}
\end{table*}

\begin{table*}[!t]
\caption{SR against different preprocessing operations. ``PT'' means ``Paint Transfer'' and ``ST'' means ``Style Transfer''.\label{tab:table3}}
\centering
\begin{tabular}{c|c|ccc|ccc|ccc|cc}
\hline\hline
\multirow{2}{*}{Task} & \multirow{2}{*}{Watermark} & \multicolumn{3}{c|}{Noise addition} & \multicolumn{3}{c|}{Resizing} & \multicolumn{3}{c|}{JPEG compression}  & \multicolumn{2}{c}{Flipping}\\ \cline{3-13} 
& & \multicolumn{1}{c|}{$\sigma =0.1$}    & \multicolumn{1}{c|}{$\sigma = 0.15$}   & $\sigma = 0.2$ & \multicolumn{1}{c|}{$128^2\times 3$} & \multicolumn{1}{c|}{$196^2\times 3$} & $512^2\times 3$ & \multicolumn{1}{c|}{QF = 50} & \multicolumn{1}{c|}{QF = 70} & \multicolumn{1}{c|}{QF = 90} & \multicolumn{1}{c|}{horizontal} & \multicolumn{1}{c}{vertical} \\ \hline
PT & \emph{Lena}  & \multicolumn{1}{c|}{100\%}   & \multicolumn{1}{c|}{100\%}   & 99.60\% & \multicolumn{1}{c|}{78.20\%} & \multicolumn{1}{c|}{98.60\%} & 100\%   & \multicolumn{1}{c|}{98.20\%} & \multicolumn{1}{c|}{98.60\%} & 99.40\% & \multicolumn{1}{c|}{100\%} & 99.00\% \\
PT & \emph{Baboon} & \multicolumn{1}{c|}{100\%}  & \multicolumn{1}{c|}{98.80\%} & 97.40\% & \multicolumn{1}{c|}{97.20\%} & \multicolumn{1}{c|}{99.80\%} & 100\%   & \multicolumn{1}{c|}{99.40\%} & \multicolumn{1}{c|}{99.60\%} & 100\%   & \multicolumn{1}{c|}{100\%} & 98.80\% \\
PT & \emph{Peppers} & \multicolumn{1}{c|}{100\%}  & \multicolumn{1}{c|}{99.40\%} & 96.60\% & \multicolumn{1}{c|}{89.20\%} & \multicolumn{1}{c|}{100\%}   & 100\%   & \multicolumn{1}{c|}{99.20\%} & \multicolumn{1}{c|}{99.40\%} & 100\%   & \multicolumn{1}{c|}{100\%} & 99.60\% \\
PT & \emph{IEEE} & \multicolumn{1}{c|}{99.60\%} & \multicolumn{1}{c|}{99.20\%} & 98.20\% & \multicolumn{1}{c|}{84.20\%} & \multicolumn{1}{c|}{99.60\%} & \multicolumn{1}{c|}{100\%}   & \multicolumn{1}{c|}{98.20\%} & \multicolumn{1}{c|}{98.40\%} & 99.40\% & \multicolumn{1}{c|}{100\%} & 99.60\% \\ \hline
ST & \emph{Lena} & \multicolumn{1}{c|}{99.73\%} & \multicolumn{1}{c|}{99.33\%} & 96.40\% & \multicolumn{1}{c|}{97.20\%} & \multicolumn{1}{c|}{99.80\%} & 99.93\% & \multicolumn{1}{c|}{93.73\%} & \multicolumn{1}{c|}{96.67\%} & 99.07\% & \multicolumn{1}{c|}{100\%} & 99.47\% \\
ST & \emph{Baboon} & \multicolumn{1}{c|}{99.47\%} & \multicolumn{1}{c|}{98.13\%} & 94.67\%  & \multicolumn{1}{c|}{83.80\%} & \multicolumn{1}{c|}{99.73\%} & 100\%  & \multicolumn{1}{c|}{95.60\%} & \multicolumn{1}{c|}{98.93\%} & 99.80\% & \multicolumn{1}{c|}{100\%} & 99.80\% \\
ST & \emph{Peppers} & \multicolumn{1}{c|}{100\%}  & \multicolumn{1}{c|}{99.53\%} & 97.33\% & \multicolumn{1}{c|}{98.07\%} & \multicolumn{1}{c|}{100\%}   & 100\%   & \multicolumn{1}{c|}{95.27\%} & \multicolumn{1}{c|}{98.07\%} & 99.00\% & \multicolumn{1}{c|}{100\%} & 100\% \\
ST & \emph{IEEE} & \multicolumn{1}{c|}{100\%}  & \multicolumn{1}{c|}{99.93\%} & 99.33\% & \multicolumn{1}{c|}{99.20\%} & \multicolumn{1}{c|}{100\%} & 100\% & \multicolumn{1}{c|}{96.00\%} & \multicolumn{1}{c|}{99.54\%} & 100\%   & \multicolumn{1}{c|}{100\%} & 100\% \\ \hline\hline
\end{tabular}
\end{table*}

\begin{table*}[!t]
    \caption{SR against the surrogate network attack, where the $\ell_1$ loss function was used for network training. \label{tab:table4}}
    \centering
    \begin{tabular}{c|c|ccc}
    \hline\hline
        Task & Watermark & ConvGen & ResGen & UnetGen \\ \hline
        Paint transfer & \emph{Lena} & 100\% & 100\% & 100\% \\
        Paint transfer & \emph{IEEE} & 99.60\% & 100\% & 99.60\% \\ \hline
        Style transfer & \emph{Lena} & 100\% & 100\% & 100\% \\
        Style transfer & \emph{IEEE} & 99.54\% & 99.80\% & 99.54\% \\ 
    \hline\hline
    \end{tabular}
\end{table*}

\begin{table*}[!t]
    \caption{SR against the surrogate network attack, where different loss functions were used for network training. \label{tab:table5}}
    \centering
    \begin{tabular}{c|c|cccccc}
    \hline\hline
    \multirow{2}{*}{Task} & \multirow{2}{*}{Watermark} & \multicolumn{6}{c}{UnetGen}\\ \cline{3-8}
    & & $\ell_1$ & $\ell_1 + \ell_\text{per}$ & $\ell_1 + \ell_\text{per} + \ell_\text{adv}$ & $\ell_2$ & $\ell_2 + \ell_\text{per}$ & $\ell_2 + \ell_\text{per} + \ell_\text{adv}$ \\ \hline
    Paint transfer & \emph{Lena} & 100\% & 100\% & 100\% & 100\% & 100\% & 100\% \\
    Paint transfer & \emph{IEEE} & 99.60\% & 100\% & 100\% & 100\% & 99.80\% & 100\% \\ \hline
    Style transfer & \emph{Lena} & 100\% & 100\% & 100\% & 100\% & 100\% & 100\% \\
    Style transfer & \emph{IEEE} & 99.54\% & 99.87\% & 99.80\% & 99.07\% & 99.67\% & 99.80\% \\ 
    \hline\hline
    \end{tabular}
\end{table*}

\subsection{Robustness}
A digital watermarking system may be attacked in realistic scenarios. It is necessary that a watermarking system is robust against common attacks. In this paper, we consider two kinds of common attacks, i.e., preprocessing and surrogate network attack. The first attack aims at manipulating the marked image so that the watermark cannot be extracted from the marked image. The second attack aims at generating a surrogate model so that the image generated by the surrogate model has good quality while revealing nothing about the watermark. We show that the proposed work can resist the two kinds of attacks.

Preprocessing is the most popular attack adopted by the attacker. Before watermark extraction, the marked image may be preprocessed by common image processing operations such as noise addition and compression. A good model watermarking technique should be able to resist these common operations. A common strategy to enhance the robustness against preprocessing is to include the preprocessed images into the training set. To this end, in the proposed framework, we utilize the attack layer $A_L$ for adversarial training to improve the robustness of model watermarking against preprocessing. 

\begin{figure}[!t]
\centering
\includegraphics[width=\linewidth]{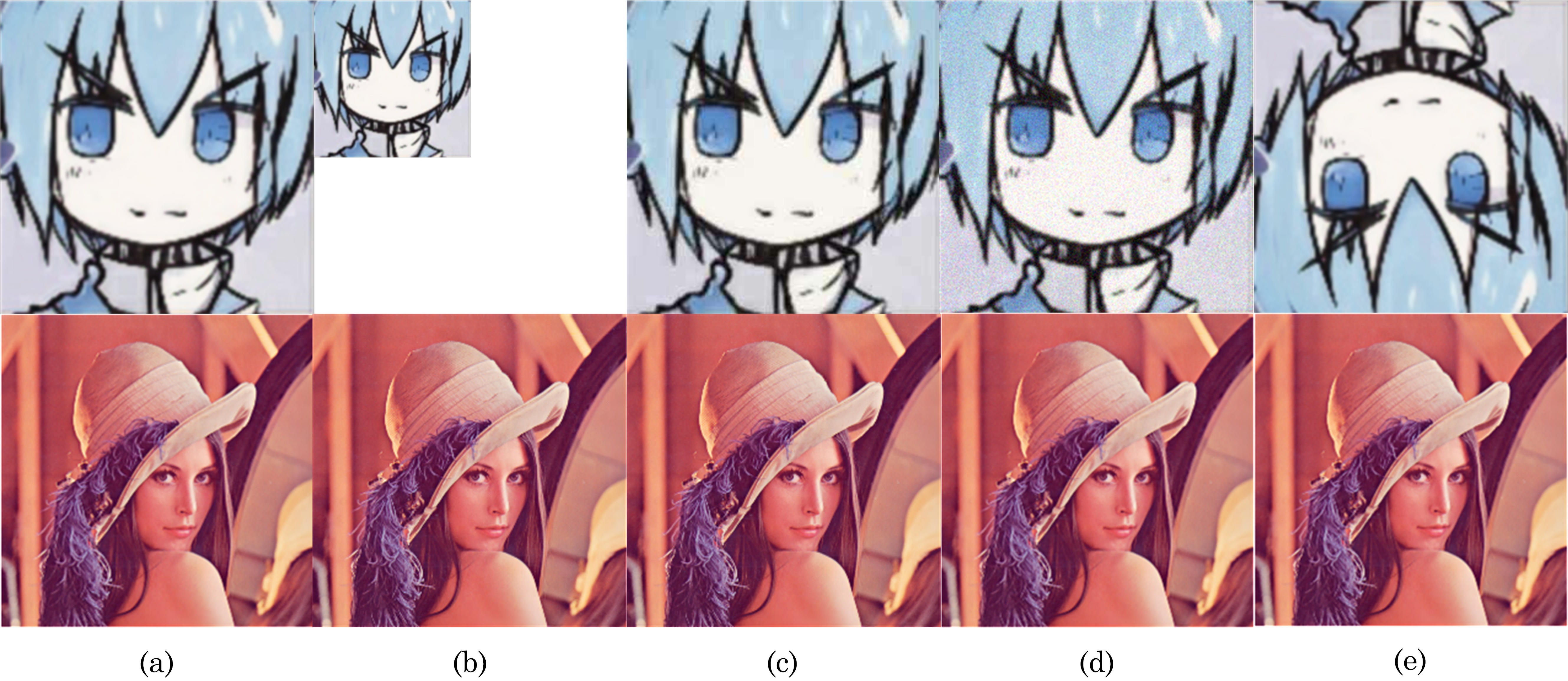}
\caption{Examples of the attacked images and extracted watermarks for paint transfer. (a) the original marked image (top) and extracted watermark (bottom), (b) the resized image ($128^2\times 3$, top) and extracted watermark (bottom), (c) the compressed image  (QF = 50, top) and extracted watermark (bottom), (d) the noised image ($\sigma$ = 0.1, top) and extracted watermark (bottom), (e) the flipped image (top) and extracted watermark (bottom).}
\label{fig_8}
\end{figure}

\begin{figure}[!t]
\centering
\includegraphics[width=\linewidth]{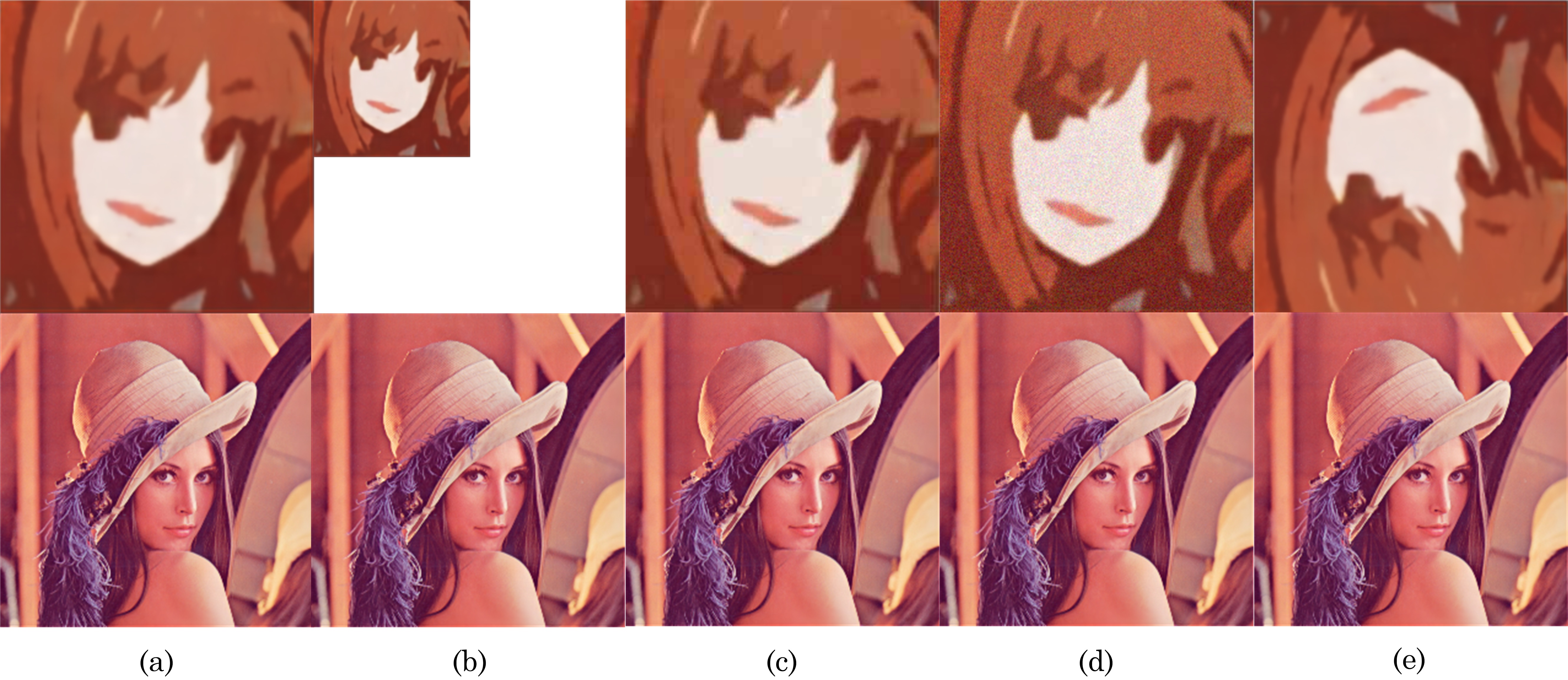}
\caption{Examples of the attacked images and extracted watermarks for style transfer. (a) the original marked image (top) and extracted watermark (bottom), (b) the resized image ($128^2\times 3$, top) and extracted watermark (bottom), (c) the compressed image  (QF = 50, top) and extracted watermark (bottom), (d) the noised image ($\sigma$ = 0.1, top) and extracted watermark (bottom), (e) the flipped image (top) and extracted watermark (bottom).}
\label{fig_9}
\end{figure}

Fig. \ref{fig_8} and Fig. \ref{fig_9} show some examples for the preprocessed images and the extracted watermarks. It can be inferred that although the marked images are subjected to different preprocessing operations, the embedded watermarks can be extracted with satisfactory quality. It indicates that adversarial training is indeed effective for improving the robustness against preprocessing. Table \ref{tab:table3} further provides the quantitative results. It can be seen that although SR tends to decline as the degree of attack increases, overall, SR can be guaranteed to be at a high level. It indicates that the proposed framework has satisfactory ability to resist preprocessing by applying adversarial training.

Surrogate network attack is a common and highly threatening attack, whose purpose is to generate a surrogate model that possesses the functionality of the original model. The surrogate model is generally obtained by training the surrogate network with the collected input-output pairs of the target host network. In order to defend against the surrogate network attack, we add a surrogate network training phase to enhance the robustness of the watermark extraction network, which can be found from Section III-C. We use different networks trained with different losses as the surrogate network to evaluate the robustness. In detail, we consider three different kinds of network structures, i.e., ``ConvGen'', ``ResGen'' and ``UnetGen''. The first network ConvGen consists of an encoder and a decoder. The encoder consists of six convolutional layers, followed by the decoder consisting of six transposed convolutional layers. The second network ResGen is the ResNet-like generative network used in CycleGAN \cite{zhu2017unpaired}. The third network UnetGen is the U-net-like network used in \cite{wu2021hiding}. The popular $\ell_1$ loss function was used to train the three networks. Table \ref{tab:table4} shows the SRs after performing the surrogate network attack. It is observed that the SRs are all close to 100\%. We can infer that the corresponding PSNRs are higher than 35 dB and most of the corresponding BERs are lower than 0.0005. As the attacker may use different loss functions, we further use different loss functions to train the surrogate network to mimic the realistic scenario. Due to the limited computational resource, we only apply the different losses to the representative network UnetGen. Table \ref{tab:table5} shows the results, in which four kinds of loss functions, i.e., $\ell_1$ loss, $\ell_2$ loss, perceptual loss $\ell_\text{per}$ \cite{johnson2016perceptual}, adversarial loss $\ell_\text{adv}$ \cite{isola2017image}, are considered. It can be seen that the proposed method allows us to successfully extract the watermark from the output of the surrogate network even though different loss functions were applied. Therefore, we can conclude that the proposed method has good ability to resist the surrogate network attack.   

\begin{table}[!t]
 \caption{Comparison between different model watermarking methods in terms of robustness and imperceptibility. \label{tab:table6}}
    \centering
    \begin{tabular}{c|c|c|c}
    \hline\hline
    \multirow{2}{*}{Method} & \multirow{2}{*}{\makecell[c]{Robustness against\\ surrogate attack}} &  \multicolumn{2}{c}{Imperceptibility}\\ \cline{3-4}
    & & Spatial domain & Frequency domain\\ \hline
    Ref. \cite{zhang2021deep} & Yes & Yes  &  \\
    Ref. \cite{wu2020watermarking} &  & Partially & \\
    Ref. \cite{zhang2022generative} & & Partially & \\
    Proposed & Yes & Yes & Yes \\ 
    \hline\hline
    \end{tabular}
\end{table}
\begin{table}[!t]

\caption{Mean PSNRs (for color watermarks, dB) and Mean BERs (for binary watermarks) before and after filtering out the high-frequency components of the marked images. ``PT'' means ``Paint Transfer'' and ``ST'' means ``Style Transfer''. The superscript ``$^*$'' means to apply the filtering operation. \label{tab:table7}}
    \centering
    \begin{tabular}{c|c|cccc}
    \hline\hline
    Task & Watermark & Ref. \cite{zhang2021deep} & Ref. \cite{wu2020watermarking} & Ref. \cite{zhang2022generative} & Proposed\\
    \hline
    PT & \emph{Lena} & 35.22 dB & 26.16 dB & 29.34 dB & 50.73 dB\\
    PT$^*$ & \emph{Lena} & 12.67 dB & 16.53 dB & 10.29 dB & 48.56 dB \\ \hline
    PT & \emph{IEEE} & 0.0027 & 0.0031 & 0.0015 & 0 \\
    PT$^*$ & \emph{IEEE} & 0.5260 & 0.4237 & 0.5726 & 0.0002 \\ \hline
    ST & \emph{Lena} & 34.05 dB & 29.48 dB & 26.01 dB & 53.74 dB\\
    ST$^*$ & \emph{Lena} & 12.65 dB & 12.93 dB & 14.10 dB & 50.04 dB\\ \hline
    ST & \emph{IEEE} & 0.0001 & 0.0004 & 0.0003 & 0 \\
    ST$^*$ & \emph{IEEE} & 0.5376& 0.4386 & 0.4515 & 0.0001 \\ 
    \hline\hline
    \end{tabular}
\end{table}

\subsection{Comparisons with Previous Methods}
We compare the proposed method with the previous methods introduced in \cite{wu2020watermarking}, \cite{zhang2021deep}, \cite{zhang2022generative}. Table \ref{tab:table6} has demonstrated the comparison results between different methods in terms of robustness against the surrogate network attack and watermark imperceptibility in the spatial/frequency domain. In Table \ref{tab:table6}, ``Yes'' means the corresponding method is robust against the corresponding attack or provides good imperceptibility in the corresponding domain. ``Partially'' means part of the marked images have good imperceptibility in the corresponding domain, i.e., some images introduce noticeable visual artifacts. It can be found that while the previous methods only achieve part of the goals, the proposed method achieves all the goals, i.e., the proposed method can resist the surrogate network attack and make the distortion caused by the secret watermark invisible in both the spatial domain and the frequency domain. We did not compare these methods in terms of robustness against preprocessing because all of them could mix the preprocessed images into the training set for resisting preprocessing.

In addition, if the adversary has the prior knowledge about frequency-domain artifacts, s/he may filter the high-frequency components of the marked image out as a way to remove the embedded watermark. To mimic such an attack, we filtered out the high-frequency components of the marked images in the test set. In detail, given a marked image, we determined the DCT coefficients of each channel of the image. Then, we kept the top-left region sized $196\times 196$ of each channel unchanged but changed the other coefficients to zero. Table \ref{tab:table7} reports the mean PSNRs/BERs of the extracted watermarks before and after filtering out the high-frequency components of the marked images for different methods. It can be found that, for all of the compared methods, after filtering, the mean PSNRs for color watermarks decline significantly, and the mean BERs for binary watermarks increase significantly. It indicates that the extracted watermarks are of poor quality. However, for the proposed method, the difference between the mean PSNRs/BERs before filtering and the mean PSNRs/BERs after filtering is very low, which indicates that the watermarks can be still extracted with high fidelity. Therefore, we can infer that compared with related works, the proposed framework has very good ability to resist high-frequency information removal.

\section{Conclusion and Discussion}
The great success brought by DNNs has also raised the need of protecting the intellectual property of advanced DNN models. In order to prevent DNN models from intellectual property infringement, many researchers apply digital watermarking to DNNs. Recent studies focus on protecting generative models, which mark the host network by marking the output such that the watermark embedded into the output can be entangled with the host network. Although these methods do not significantly impair the original task of the host network during watermark embedding and enable the watermark to be extracted for intellectual property protection, they introduce noticeable visual artifacts in the high-frequency domain, which is mainly caused by the sampling operation applied in the network architecture, and therefore reduces the imperceptibility of the watermark. It may lead the adversary to reveal the presence of the watermark and further remove the embedded watermark, which threatens the intellectual property protection of DNN models. To address this problem, in this paper, we propose a general framework for generative model watermarking that can suppress the high-frequency artifacts. The main idea is to exploit anti-aliasing for the design of the watermark embedding network. Meanwhile, joint loss optimization and adversarial training are applied to further enhance the effectiveness and robustness. Experiments indicate that our method not only maintains the performance very well on the original task, but also demonstrates better imperceptibility and robustness on watermarking compared with related works, which verify the superiority and applicability.

We should admit that we cannot guarantee that the proposed framework is robust against all the real-world attacks since it is impossible for us to foresee all the attacks performed by the adversary. Actually, even for existing methods, they only resist specific attacks. However, by removing the artifacts caused by watermark embedding, the watermark can be concealed very well, which significantly improves the imperceptibility of the watermark and may not arouse suspicion of the adversary. In other words, by improving the imperceptibility, the probability of the watermark being attacked can be significantly reduced, thereby improving the reliability of intellectual property protection of DNN models. We hope this attempt can inspire more advanced works in the future.

\bibliographystyle{IEEEtran}
\bibliography{reference.bib}

\vfill

\end{document}